\documentclass{llncs}

\usepackage[T1]{fontenc}
\usepackage{amsmath,amssymb,amsfonts}
\usepackage{graphicx}
\usepackage{float}
\usepackage{booktabs}
\usepackage{xcolor}
\usepackage{bm}
\usepackage{mathtools}
\usepackage{hyperref}
\usepackage{url}

\hypersetup{colorlinks=true, linkcolor=blue, citecolor=blue, urlcolor=blue}

\newcommand{\R}{\mathbb{R}}
\newcommand{\C}{\mathbb{C}}
\newcommand{\norm}[1]{\left\lVert #1 \right\rVert}
\newcommand{\statevec}[1]{\left| #1 \right\rangle}
\newcommand{\braket}[2]{\left\langle #1 \mid #2 \right\rangle}
\newcommand{\cond}{\kappa}
\newcommand{\HilbertH}{\mathcal{H}}
\DeclareMathOperator*{\argmin}{arg\,min}

\begin{document}

\title{A Quantum Reservoir Architecture for Chaotic Forecasting
and a Test of Whether Its High Dimension Helps}

\titlerunning{A Quantum Reservoir Architecture for Chaotic Forecasting}

\author{Tushar Pandey}
\authorrunning{T. Pandey}
\institute{Texas A\&M University, College Station, TX, USA}

\maketitle

\begin{abstract}
Quantum reservoir computing uses a fixed quantum circuit as a feature
generator and trains only a simple linear readout on top of it. This makes
it cheap to train and free of the optimisation problems that affect many
quantum machine-learning models. A natural worry is that the very large
feature space the circuit produces might inflate apparent performance
without adding anything real. This paper provides two things. First, it
gives a complete, reproducible recipe for one such reservoir applied to
forecasting chaotic systems, including how data is fed in, how the circuit
is built, and how the readout is trained. Second, it gives a way to tell
whether the reservoir's high dimension is actually doing useful work. We
grow the size of the prediction problem and the size of the quantum
reservoir together, so that extra capacity cannot be the explanation for
any improvement, and we track a single stability number that measures how
well behaved the readout fit is. On two chaotic test systems, a
spatiotemporal chain and a shallow-water fluid model, the quantum
reservoir keeps a flat, stable error as both sizes grow, while a matched
classical reservoir does not. We report where the classical baseline is in
fact stronger, so the comparison is honest. The result is a clean
specification plus a diagnostic that other groups can apply to any
reservoir whose features have a known scale.
\end{abstract}

\section{Introduction}

Reservoir computing replaces the gradient-trained recurrent network with a
fixed nonlinear dynamical system whose state is read out by a trainable
linear map. The two design choices that survive in such a model are (i) the
structure of the reservoir, which determines the feature map
$x \mapsto \phi(x)$, and (ii) the readout, which is a ridge-regularised
linear regression. A quantum reservoir replaces the classical nonlinear
recurrence with a fixed quantum circuit and uses basis-state probabilities
as the feature vector \cite{fujii2017,nakajima2019boosting,mujal2021}.

For chaotic dynamical systems the appeal of a reservoir is operational. No
quantum gradient is required, no barren-plateau pathologies arise from
circuit depth \cite{mcclean2018}, and the cost of training is the cost of
one ridge solve. The corresponding question is architectural. Which
encoding, which entangling pattern, and which readout structure produce a
feature map whose statistical properties remain controlled as either the
qubit count $q$ or the prediction dimension $K$ grows?

This paper is a methodological reference. We give a gate-level
specification of the quantum reservoir, describe the two pipelines in which
we deploy it (autonomous prediction on proper-orthogonal-decomposition
coefficients, and additive correction of a deliberately crude physics
core), and develop a scaling protocol in which $K$ and $q$ are increased
together. We then state the architectural argument that motivates the
design. Under angle encoding into a fixed entangling circuit, the readout
features are basis-state probabilities
$p_s = \lvert\braket{s}{\psi}\rvert^2$, supported on the probability
simplex. The resulting feature Gram matrix admits a spectral bound that is
independent of $q$, which we conjecture underlies the empirically flat
scaling of the model's prediction error. The empirical results, replicated
on two independent dynamical systems across their matched-complexity sweeps,
are reported in a single condensed section. The bulk of the paper is the
architecture and the scaling protocol.

\paragraph{Relation to prior work.}
This paper is a methods follow-up to the author's earlier benchmark study
\cite{pandey2026qrc}, which compares a fixed quantum reservoir against a
variational quantum physics-informed network (QPINN) on the Lorenz system
and asks which paradigm forecasts chaos more reliably. That work is a
head-to-head benchmark of two quantum architectures. The present paper does
not re-run that benchmark. Its contributions are distinct and
complementary: (i) a complete gate-level architecture specification for the
quantum reservoir together with two deployment pipelines, (ii) a
conditioning result showing that the basis-state-probability feature Gram
of the quantum reservoir is bounded in condition number independently of
qubit count, while the matched classical baseline degrades, and (iii) a
matched-complexity ($K = q$) scaling protocol that grows the prediction
target and the reservoir together so that improvements cannot be attributed
to spare capacity. In short, \cite{pandey2026qrc} establishes that the
fixed quantum reservoir is competitive; this paper specifies the
architecture precisely and explains, through conditioning, why its
behaviour is stable as it scales.

\section{Background and notation}

\paragraph{Notation.}
We write $x(t) \in \R^N$ for the state of an $N$-dimensional dynamical
system and $a(t) \in \R^K$ for a $K$-dimensional reduced state obtained by
linear projection from $x(t)$ (e.g.\ POD coefficients). Lowercase Greek
letters denote angles or hyperparameters; $q$ denotes the qubit count and
$\HilbertH = (\C^2)^{\otimes q}$ the Hilbert space of $q$ qubits with
computational basis $\{\statevec{s}\}_{s \in \{0,1\}^q}$.

\paragraph{Echo state networks.}
A classical echo-state network (ESN) maintains a hidden state
$r_t \in \R^{N_{\mathrm{res}}}$ updated by
$r_t = (1-\alpha_{\mathrm{leak}})\, r_{t-1}
+ \alpha_{\mathrm{leak}}\,\tanh(W_{\mathrm{res}}\, r_{t-1}
+ W_{\mathrm{in}}\, x_t)$,
with $W_{\mathrm{res}}$ sparse with prescribed spectral radius and
$W_{\mathrm{in}}$ random with prescribed input scaling
\cite{jaeger2004harnessing,pathak2018model}. The features $r_t$ are read out
by a linear map trained by ridge regression. We use the ESN as the
matched-feature classical baseline throughout.

\paragraph{Quantum reservoir computing.}
The quantum analogue replaces the classical recurrence by a fixed
parametrised circuit $U_{\mathrm{res}}$. Given a real input $x_t$, an
encoding circuit $U_{\mathrm{enc}}(x_t)$ prepares
$\statevec{\psi(x_t)} = U_{\mathrm{res}}\, U_{\mathrm{enc}}(x_t)\,
\statevec{0}^{\otimes q}$, and the feature vector is the vector of
computational-basis probabilities,
\[
\phi_q(x_t) \coloneqq
\bigl(\,\lvert\braket{s}{\psi(x_t)}\rvert^2\,\bigr)_{s \in \{0,1\}^q}
\in \Delta^{2^q-1},
\]
where $\Delta^{d-1}$ denotes the $(d-1)$-simplex. Two structural properties
follow immediately: $\phi_q(x_t) \ge 0$ componentwise, and
$\sum_s \phi_q(x_t)_s = 1$ for every $x_t$.

\paragraph{POD reduction for spatial fields.}
For a 2D PDE state $u(\mathbf{x}, t)$ sampled on a grid we collect training
snapshots into a matrix $X \in \R^{T \times M}$ with $M$ the total number of
grid degrees of freedom, normalise each physical component by its empirical
standard deviation, subtract the temporal mean, and take the singular value
decomposition $X_c = U \Sigma V^\top$. The first $K$ right singular vectors
form the POD basis $\Phi \in \R^{M \times K}$, and the reduced state at time
$t$ is $a(t) = X_c(t)\, \Phi \in \R^K$
\cite{cordier2010pod,ahmed2021closures}. The reconstruction is the linear
lift $\hat x(t) = a(t)\,\Phi^\top + \bar x$, with $\bar x$ the training
mean.

\section{The quantum reservoir architecture}
\label{sec:architecture}

The trainable-free quantum stage consists of an encoding block followed by
a fixed reservoir block; the trainable stage is a temporal-windowed ridge
readout. Inputs are angle-encoded onto $q$ qubits by one $R_Y$ rotation per
qubit, then passed through $L = 2$ fixed layers, each a block of random
single-qubit rotations followed by a data-dependent CNOT--$R_Z$ entangling
chain. Distinct inputs therefore traverse distinct unitaries, which is what
gives the circuit its reservoir-like behaviour; only the random rotation
angles are fixed by a single seed. The state is read out in the
computational basis under exact statevector evaluation, giving the
deterministic feature vector
$\phi_q(x_t) = (\lvert\braket{s}{\psi(x_t)}\rvert^2)_{s \in \{0,1\}^q}
\in \Delta^{2^q-1}$. To supply trajectory history the last $w = 5$ feature
vectors are concatenated into a Takens-style delay embedding
\cite{takens1981} $F_t \in \R^{w 2^q}$, and the readout is the closed-form
ridge regression
\begin{equation}
W^\star = \argmin_{W} \norm{F W - S}_F^2 + \alpha\, \norm{W}_F^2
= (F^\top F + \alpha I)^{-1} F^\top S,
\label{eq:ridge}
\end{equation}
with $\alpha > 0$ fixed. Training is one linear solve, under
$0.2\,\mathrm{s}$ for all configurations. We deploy the reservoir in two
pipelines: \emph{Block~A (autonomous prediction)} maps the reduced state
$a(t)$ directly to $a(t+1)$ (or its increment) with no physics solver in the
loop, the natural setting to compare against the matched classical reservoir
at the same feature dimension; \emph{Block~B (additive physics correction)}
runs a deliberately crude physics integrator and trains the reservoir on the
per-step POD-space residual, connecting to hybrid neural-physical
forecasting \cite{kochkov2024neuralgcm,lam2023graphcast}. The full
gate-level specification, including the encoding angles, the entangling
block, the measurement model, the state/delta targets and teacher-forced
versus closed-loop evaluation, and the Block~B residual lift, is given in
Appendix~\ref{app:architecture}.

\section{Matched-complexity scaling design}
\label{sec:scaling}

The quantum reservoir's feature dimension is $2^q$, which grows
exponentially in qubit count. A natural diagnostic question is whether this
growth translates into expressive power useful for prediction. The standard
sweep, fix the problem and scale $q$, conflates two distinct effects.

Let $K$ denote the dimensionality of the reduced prediction target ($K = N$
for L96; $K =$ POD rank for SWE), and let $q$ denote the qubit count. The
matched-complexity sweep prescribes $K = q$, so the regression target grows
in lockstep with the feature dimension. This design is diagnostic. Under
$K = q$, the per-feature capacity of the reservoir relative to the
regression problem is approximately constant: every doubling of $q$ doubles
the target dimension, so the readout must allocate the same budget of
features per prediction component. A method whose error remains flat across
the sweep is therefore not exploiting growing spare capacity. It is
exhibiting an architectural property that survives the natural ``you just
gave it more parameters'' rebuttal. By contrast, the design that fixes $K$
and sweeps $q$ measures the marginal value of additional features for the
same task and is dominated by the problem dimension rather than by any
architectural property.

Each cell in a sweep is run with $N_{\mathrm{seeds}} = 8$ independent seeds;
the seed controls the reservoir's random rotation angles (and the ESN's
random matrices). Within a cell we report mean and standard deviation of
the test MSE, the per-seed win count of the proposed method against each
baseline, and the paired Wilcoxon signed-rank $p$-value. The Wilcoxon test
is appropriate because seeds are paired, the differences are not
necessarily Gaussian, and the statistic depends only on the sign of the
per-seed difference. We additionally record a directional audit: a
disagreement between the per-seed-mean verdict and the seed-win-count
verdict flags an outlier-driven mean difference.

\section{The conditioning argument}
\label{sec:conditioning}

The architectural property that motivates the $K = q$ design is the
following. The features $\phi_q(x_t)$ are basis-state probabilities
constrained to the simplex; the corresponding ridge solve has a spectral
structure that admits a $q$-independent bound.

\paragraph{Spectral bound for QRC features.}
For a single feature vector, $\phi_q(x_t) \in \Delta^{2^q-1}$ implies
$\norm{\phi_q(x_t)}_2 \le 1$. For the windowed feature
$F_t \in \R^{w 2^q}$ we therefore have $\norm{F_t}_2 \le \sqrt{w}$. The Gram
matrix of the training feature matrix
$F \in \R^{(T-w+1) \times w 2^q}$ thus satisfies
$\norm{F^\top F}_2 \le \norm{F}_F^2 \le (T-w+1)\, w$, independently of $q$.
For ridge regression with regulariser $\alpha$,
\begin{equation}
\cond(F^\top F + \alpha I)
= \frac{\sigma_{\max}^2(F) + \alpha}{\sigma_{\min}^2(F) + \alpha}
\le \frac{(T-w+1)\, w + \alpha}{\alpha},
\label{eq:cond-bound}
\end{equation}
which is $q$-independent. The bound is loose when many training features are
nearly orthogonal, in which case $\sigma_{\min}^2(F)$ is itself
$\Theta(1)$.

\paragraph{Comparison to ESN features.}
ESN features $r_t \in (-1,+1)^{N_{\mathrm{res}}}$ are bounded componentwise
but not normalised in $\ell_2$. With $N_{\mathrm{res}} = 2^q$ units,
$\norm{r_t}_2$ can be as large as $2^{q/2}$, so the Gram spectrum is bounded
only by $(T-w+1)\, w\, 2^q$. The empirical condition number grows with $q$
because the dense random recurrent matrix produces a heavy-tailed singular
value distribution amplified at higher state dimension. Ridge regression on
such a Gram matrix overfits non-monotonically across seeds, producing the
oscillating test MSE observed in Section~\ref{sec:results}.

\paragraph{Falsifiable prediction.}
The argument predicts that $\cond(F_{\mathrm{QRC}}^\top F_{\mathrm{QRC}} +
\alpha I)$ remains bounded as $K = q$ grows, while
$\cond(F_{\mathrm{ESN}}^\top F_{\mathrm{ESN}} + \alpha I)$ grows. Direct
measurement of these condition numbers from the same $(K, \mathrm{seed})$
cells used for the test-MSE evaluation gives a single-number diagnostic that
confirms or falsifies the prediction; we report it in
Section~\ref{sec:results}.

\section{Two test systems}
\label{sec:systems}

\paragraph{Lorenz-96.}
The Lorenz-96 model \cite{lorenz1996} with forcing $F = 8$ is
$\dot x_i = (x_{i+1} - x_{i-2})\, x_{i-1} - x_i + F$, $i = 0, \dots, N-1$,
with periodic indexing. Each $x_i$ is a genuine degree of freedom with no
reduced-order bookkeeping, so setting $q = N$ gives a one-to-one assignment
of qubits to variables and the map $\pi$ in
eq.~\eqref{eq:angle-encoding} is the identity. We integrate with RK4 at
$\Delta t = 0.01$, train on 50 evenly-spaced subsamples of $[0, 3]$,
evaluate on 20 subsamples of $[3, 4]$, and sweep
$q = N \in \{5\text{--}11\}$.

\paragraph{Doubly-periodic shallow water.}
The rotating shallow-water equations on a doubly-periodic $L \times L$
domain in vector-invariant form are
$\partial_t u = q V - \partial_x (K + g h)$,
$\partial_t v = -q U - \partial_y (K + g h)$,
$\partial_t h = -\partial_x U - \partial_y V$,
with mass fluxes $U = u h$, $V = v h$, kinetic energy $K = (u^2 + v^2)/2$,
and potential vorticity $q = (\zeta + f)/h$. Integration uses a Sadourny
enstrophy-conserving Arakawa C-grid scheme with biharmonic dissipation. The
full nonlinear solver gives the ground truth; a linearised variant is the
crude physics core $\mathcal{P}$ for Block~B. The grid is $64 \times 64$,
$L = 10^6\,\mathrm{m}$, $g = 9.81\,\mathrm{m\,s^{-2}}$,
$H = 100\,\mathrm{m}$, $f_0 = 10^{-4}\,\mathrm{s^{-1}}$. POD is fit on
component-wise standardised training snapshots, and we sweep
$q = K \in \{5\text{--}8\}$ POD modes per cell.

\section{Results}
\label{sec:results}

We summarise the empirical evidence in a single section.

\begin{figure}[H]
\centering
\includegraphics[width=0.62\linewidth]{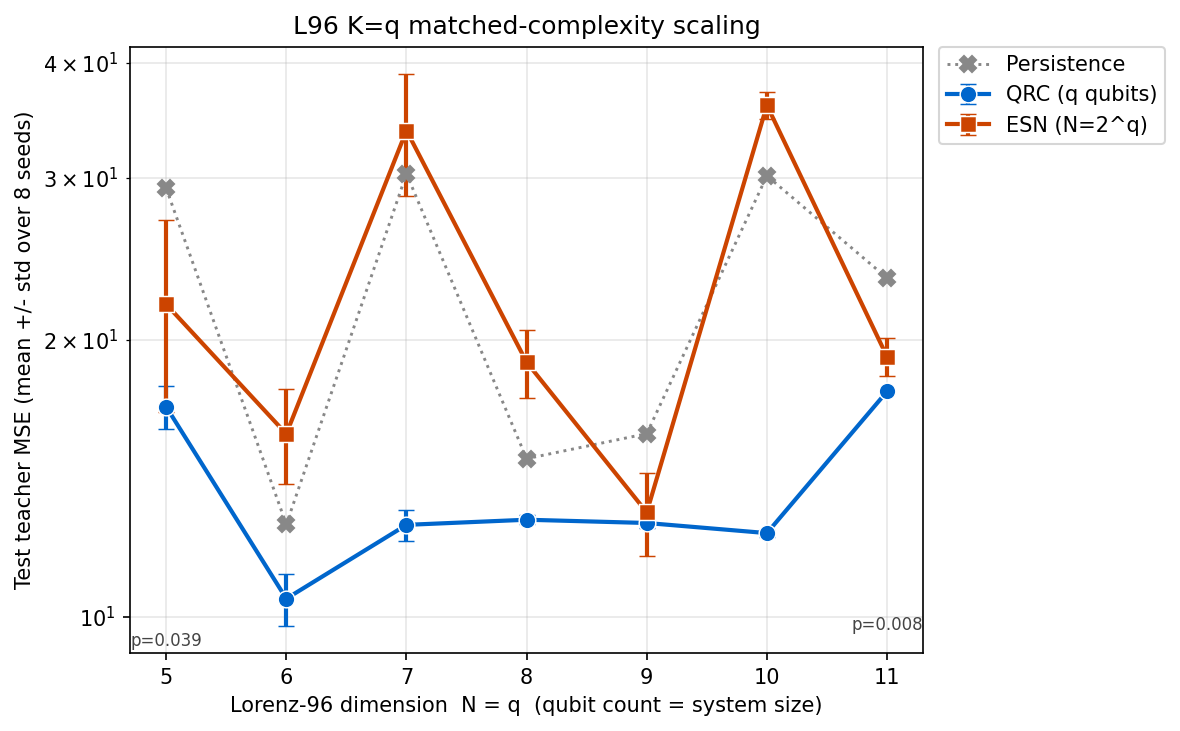}
\caption{Test teacher MSE for QRC ($q$ qubits) and matched classical ESN
($N_{\mathrm{res}} = 2^q$) on Lorenz-96 with $q = N \in \{5\text{--}11\}$.
Mean $\pm$ std over 8 seeds per cell. QRC remains essentially flat; the ESN
oscillates non-monotonically with feature dimension.}
\label{fig:l96-kq}
\end{figure}

\paragraph{L96 matched-complexity scaling.}
QRC's mean test MSE is approximately constant across the sweep, while the
matched-feature ESN ($N_{\mathrm{res}} = 2^q$) oscillates non-monotonically
with $q$ (Fig.~\ref{fig:l96-kq}). Five of seven cells reach $p \le 0.039$ on
the paired Wilcoxon test, with the leftmost cell at $p = 0.008$ (8/8 seed
wins). One cell ($N = 9$) ties at $p = 0.64$ and is bracketed by significant
QRC wins on either side, consistent with the non-monotonic conditioning
behaviour predicted in Section~\ref{sec:conditioning}.

\paragraph{SWE matched-complexity scaling and hybrid forecast.}
On SWE Block~A the QRC-vs-ESN comparison yields
$p \in \{0.016, 0.008, 0.008, 0.008\}$ across $K = q \in \{5\text{--}8\}$
(per-comparison $p$-values plotted in Fig.~\ref{fig:kq-swe},
Appendix~\ref{app:architecture}). On Block~B the lead-1 hybrid achieves
significant improvement over the physics-only baseline at all four cells. We
note an asymmetry important for honest interpretation: at lead-1 in Block~B
the matched-feature ESN is a \emph{stronger} corrector than QRC at low $K$,
with the gap closing as $K$ grows.

\begin{figure}[H]
\centering
\includegraphics[width=0.62\linewidth]{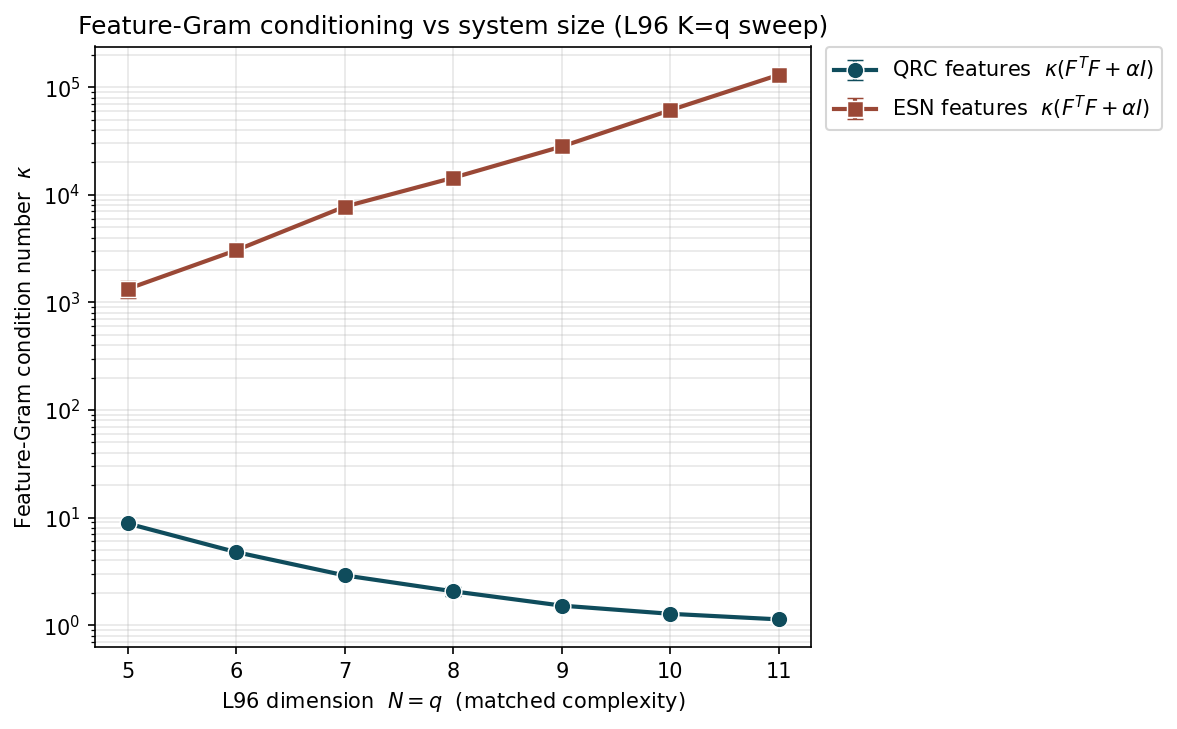}
\caption{Feature-Gram condition number $\cond(F^\top F + \alpha I)$ vs L96
dimension $N = q$ for QRC and matched-feature ESN, 8 seeds per cell. QRC
conditioning decreases toward $1$; ESN conditioning grows monotonically.
The ratio reaches $\sim 5 \times 10^4$ at $N = 10$.}
\label{fig:conditioning}
\end{figure}

\paragraph{Conditioning measurements.}
We directly measure the empirical condition number
$\cond(F^\top F + \alpha I)$ at each $(K, \mathrm{seed})$ cell for both
feature matrices, on the same training-set features used in the ridge solve.
The measurement (Fig.~\ref{fig:conditioning}; per-cell values in
Table~\ref{tab:conditioning}, Appendix~\ref{app:architecture})
confirms the falsifiable prediction on both sides.
$\cond_{\mathrm{QRC}}$ decreases monotonically from $8.8$ at $N = 5$ to
$1.13$ at $N = 11$, approaching the perfect-conditioning floor $\cond = 1$
from above, so the bound eq.~\eqref{eq:cond-bound} is increasingly
saturated as $N$ grows. $\cond_{\mathrm{ESN}}$ grows by roughly a factor of
$2$ per unit increase in $N$, reaching $1.3 \times 10^5$ at $N = 11$, so
the ratio $\cond_{\mathrm{ESN}} / \cond_{\mathrm{QRC}}$ exceeds $10^5$. The conditioning argument thus moves
from conjecture to direct measurement. We retain the caveat
(Section~\ref{sec:discussion}) that conditioning is distinct from the bias
of the ridge estimator.

\section{Discussion}
\label{sec:discussion}

\paragraph{What the architecture commits to.}
The reservoir commits to three substantive choices: angle encoding into
single-qubit $R_Y$ gates, two layers of random single-qubit rotations
followed by a data-dependent CNOT--$R_Z$ entangling chain, and a
temporal-windowed ridge readout on basis-state probabilities. Each is
replaceable in principle. Amplitude encoding, IQP-style ansatz,
hardware-native gate sets, expectation-value features, and recurrent readout
structures are all consistent with the broader programme. We do not claim
optimality, only that the architecture is fully specified and that the
matched-complexity behaviour is reproducible from a single random seed.

\paragraph{Where the conditioning argument bites.}
The bound eq.~\eqref{eq:cond-bound} controls the conditioning of the ridge
solve, not the bias of the ridge estimator. A well-conditioned problem may
still predict poorly if the features are uninformative. The argument
therefore explains the \emph{stability} of QRC's test MSE across
$(K, \mathrm{seed})$, the absence of the ESN's non-monotonic oscillation,
but does not explain the \emph{level} of QRC's error relative to other
classical methods. A broader baseline panel (linear regression, random
Fourier features at matched dimension, a trainable MLP at matched parameter
count) is needed to disentangle these effects and is ongoing work. We have
additionally observed, in a preliminary L96 sweep at $N = 8$ varying the
forcing $F$, that the QRC--ESN gap reverses sign across the chaos onset:
ESN dominates at low forcing where the system is near-stable, while QRC
overtakes it beyond $\lambda_{\max} \approx 1$. This is the
conditioning-failure mode predicted here and is the subject of separate work
in preparation.

\paragraph{Limitations.}
(i) \emph{Block~B at lead-1.} The matched-feature ESN beats QRC at low $K$
in the hybrid pipeline; the QRC advantage there is restricted to the
rollout-mean over many lead steps. (ii) \emph{Simulator scale.} All
experiments use exact statevector evaluation and are noiseless. Sampling
noise from finite shots and hardware decoherence will degrade
$\phi_q(x_t)$ in ways the conditioning argument does not address. (iii)
\emph{Baseline panel.} The current baseline is the matched-feature ESN and a
persistence floor; adding the broader panel above is in progress. (iv)
\emph{Hyperparameter scope.} We varied $(\alpha, w, L)$ in a single-cell
sensitivity sweep; the choices $(1.0, 5, 2)$ are reasonable but not defended
as globally optimal.

\section{Conclusions}

We have given a complete gate-level specification of a fixed-Hamiltonian
quantum reservoir for chaotic-system forecasting and a matched-complexity
scaling protocol that grows the qubit count and the prediction-target
dimension together. Training is one ridge solve, with no quantum gradient
and no barren-plateau pathology. The protocol gives a diagnostic not
explainable by extra features, and the argument behind it, the bounded
conditioning of the basis-state-probability feature Gram, is verified
empirically on two dynamical systems across their matched-complexity sweeps. We
have been explicit about where the classical baseline is stronger and about
the limitations that remain. The methodological contribution stands
independently: the architecture is fully specified, the protocol is
portable, and the conditioning argument is testable on any reservoir with a
known feature normalisation.

\subsubsection*{Code and data availability.}
All code, data, and scripts to reproduce the figures and tables are
available in the public repository
\url{https://github.com/pandey-tushar/Quantum_Chaos_solver} on the
\texttt{paper-2-methods} branch.

\subsubsection*{Acknowledgements.}
The author thanks the broader quantum reservoir computing community for
discussions on fixed-Hamiltonian reservoir architectures. The methodological
framing was informed by prior results on QRC for discrete chaotic maps.

\bibliographystyle{splncs04}
\bibliography{refs}

\appendix
\section{Full gate-level architecture and supporting measurements}
\label{app:architecture}

This appendix gives the complete gate-level specification summarised in
Section~\ref{sec:architecture}, together with the per-cell conditioning table
and the SWE significance figure.

\paragraph{Angle encoding.}
Given an input $x_t \in \R^N$, define the per-component angles
$\varphi_t \in [0, 2\pi]^q$ by
\begin{equation}
\varphi_{t,i} =
\frac{x_{t,\,\pi(i)} - \ell_i}{u_i - \ell_i + \varepsilon}\, \cdot\, 2\pi,
\qquad i = 0,\dots,q-1,
\label{eq:angle-encoding}
\end{equation}
where $\pi: \{0,\dots,q-1\} \to \{0,\dots,N-1\}$ assigns input components to
qubits and $(\ell_i, u_i)$ are component-wise bounds. We use a global
convention, where $(\ell_i, u_i)$ are the training-set minimum and maximum
of component $\pi(i)$ fixed at training time, to avoid data leakage; for
L96, which is statistically stationary, a per-state convention
($\ell = \min_j x_{t,j}$, $u = \max_j x_{t,j}$) is also admissible. The
encoding circuit applies one rotation gate per qubit,
$U_{\mathrm{enc}}(x_t) = \prod_{i=0}^{q-1} R_Y(\varphi_{t,i})_i$. In the
matched-complexity setting $q = K$ (Section~\ref{sec:scaling}) the
assignment $\pi$ is a bijection.

\paragraph{The fixed reservoir circuit.}
The reservoir circuit is composed of $L$ identical layers, each a
single-qubit rotation block followed by an entangling block. We use $L = 2$
throughout. At layer $\ell$, qubit $i$ undergoes
$R_X(\theta^{(\ell)}_{i,1})\, R_Y(\theta^{(\ell)}_{i,2})\,
R_Z(\theta^{(\ell)}_{i,3})$, with angles drawn from
$\mathrm{Uniform}[0, 2\pi)$ at initialisation and fixed thereafter (set by a
single random seed). The entangling block is a CNOT chain along the qubit
ordering with a phase rotation after each gate,
\begin{equation}
\prod_{i=0}^{q-2} \bigl[ R_Z(c\,\varphi_{t,i})_{i+1}
\cdot \mathrm{CNOT}_{i \to i+1} \bigr],
\label{eq:ent-block}
\end{equation}
optionally closed into a ring. The coupling strength $c = 0.5$ is common to
all phase rotations. The entangling phase $R_Z(c\,\varphi_{t,i})$ depends on
the input, so distinct inputs traverse distinct unitaries; only the random
rotation angles are fixed. This data dependence is what gives the circuit
its reservoir-like behaviour.

\paragraph{Measurement and feature vector.}
After $L$ layers the state $\statevec{\psi(x_t)}$ is measured in the
computational basis. We use exact statevector evaluation, so the feature
vector is the deterministic
$\phi_q(x_t) = (\lvert\braket{s}{\psi(x_t)}\rvert^2)_{s \in \{0,1\}^q}
\in \Delta^{2^q - 1}$.
Hardware deployment would replace this with a Monte-Carlo estimate from
$N_{\mathrm{shots}}$ repetitions; the resulting sampling noise is one of the
limitations discussed in Section~\ref{sec:discussion}.

\paragraph{Temporal windowing and ridge readout.}
A single feature vector contains no trajectory history, so we concatenate
the last $w$ feature vectors,
\begin{equation}
F_t = \bigl[\phi_q(x_{t-w+1}) \,\Vert\, \cdots \,\Vert\, \phi_q(x_t)\bigr]
\in \R^{w 2^q},
\end{equation}
a Takens-style delay embedding \cite{takens1981} in the lifted feature
space ($w = 5$ throughout). Let $\mathcal{D} = \{(F_t, s_t)\}_{t=w}^{T}$ be
the training set with $s_t \in \R^d$ the target; stacking rows gives
$F$ and $S$. The readout is the closed-form ridge regression of
eq.~\eqref{eq:ridge}, with $\alpha > 0$ fixed and $\norm{\cdot}_F$ the
Frobenius norm. Training cost is dominated by forming $F^\top F$ and solving
the linear system; in our experiments this is under $0.2\,\mathrm{s}$ for all
configurations.

\paragraph{Targets and prediction modes.}
Two targets are used: \emph{state} ($s_t = a(t+1)$, used for L96) and
\emph{delta} ($s_t = a(t+1) - a(t)$, used for SWE where the reduced
coefficients have sharply varying dynamic range and the increment is the
better-conditioned target). At evaluation we report a \emph{teacher-forced}
mode (features from ground-truth states, isolating the single-step map) and
a \emph{closed-loop} mode (features from the model's own outputs after a
burn-in of length $w$, assessing rollout stability).

\paragraph{Block A and Block B pipelines.}
\emph{Block A (autonomous prediction)} maps the reduced state $a(t)$
directly to $a(t+1)$ (or its increment) with no physics solver in the loop;
this is the natural setting to compare the quantum reservoir against the
matched classical reservoir at the same feature dimension. \emph{Block B
(additive physics correction)} runs a deliberately crude physics integrator
$\mathcal{P}$ on the full grid state to produce
$x_{\mathcal{P}}(t+1) = \mathcal{P}(x(t))$, and trains the reservoir on the
per-step residual in POD space,
$r(t) = \Phi^\top[\, x(t+1) - x_{\mathcal{P}}(t+1) \,] \in \R^K$. The
corrected next-state $\hat x(t+1) = x_{\mathcal{P}}(t+1) + \hat r(t)\,
\Phi^\top$ is applied in grid space to avoid mode-truncation instabilities
at long horizons. This residual-correction setting connects to hybrid
neural-physical forecasting \cite{kochkov2024neuralgcm,lam2023graphcast}.

\begin{table}[H]
\centering
\caption{Measured feature-Gram condition number across the L96 $K = q$
sweep (8 seeds, mean $\pm$ std, $\alpha = 1.0$).}
\label{tab:conditioning}
\setlength{\tabcolsep}{9pt}\renewcommand{\arraystretch}{1.15}
\begin{tabular}{cccr}
\toprule
$N$ & $\cond_{\mathrm{QRC}}$ & $\cond_{\mathrm{ESN}}$ &
ratio $\cond_{\mathrm{ESN}} / \cond_{\mathrm{QRC}}$ \\
\midrule
$5$  & $8.81 \pm 0.49$  & $1.33 \times 10^3 \pm 2.4 \times 10^2$ & $1.5 \times 10^2$ \\
$6$  & $4.77 \pm 0.14$  & $3.05 \times 10^3 \pm 2.5 \times 10^2$ & $6.4 \times 10^2$ \\
$7$  & $2.90 \pm 0.07$  & $7.76 \times 10^3 \pm 4.6 \times 10^2$ & $2.7 \times 10^3$ \\
$8$  & $2.06 \pm 0.18$  & $1.44 \times 10^4 \pm 8.1 \times 10^2$ & $7.0 \times 10^3$ \\
$9$  & $1.52 \pm 0.04$  & $2.81 \times 10^4 \pm 1.1 \times 10^3$ & $1.9 \times 10^4$ \\
$10$ & $1.28 \pm 0.02$  & $6.11 \times 10^4 \pm 2.3 \times 10^3$ & $4.8 \times 10^4$ \\
$11$ & $1.13 \pm 0.01$  & $1.30 \times 10^5 \pm 2.8 \times 10^3$ & $1.2 \times 10^5$ \\
\bottomrule
\end{tabular}
\end{table}

\begin{figure}[H]
\centering
\includegraphics[width=0.62\linewidth]{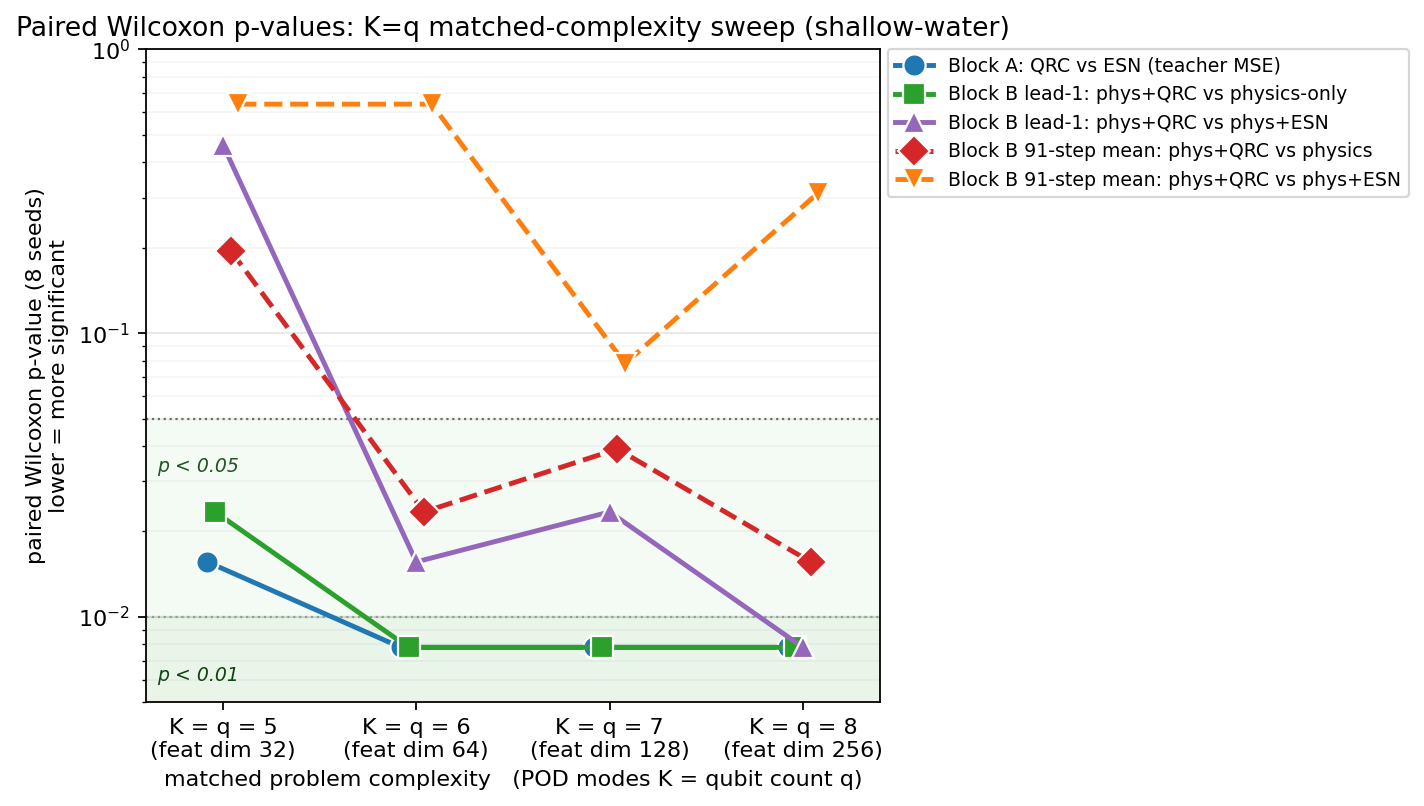}
\caption{Paired Wilcoxon $p$-values for the matched-complexity sweep on SWE
Block~A and Block~B. Four of the five comparisons enter the significant
region ($p < 0.05$) as $K = q$ grows; the fifth (91-step-mean
physics+QRC vs.\ physics+ESN) does not.}
\label{fig:kq-swe}
\end{figure}

\end{document}